# Computer Systems to Oil Pipeline Transporting

**Assoc.Prof. Timur Chis, Ph.D., Dipl.Eng.**
**"Andrei Saguna" University, Constanta, Romania**

ABSTRACT. Computer systems in the pipeline oil transporting that the greatest amount of data can be gathered, analyzed and acted upon in the shortest amount of time. Most operators now have some form of computer based monitoring system employing either commercially available or custom developed software to run the system. This paper presented the SCADA systems to oil pipeline in concordance to the Romanian environmental reglementations.

## 1 Introduction

Computer systems in the Romanian Oil Pipeline are now technology based to SCADA theory. SCADA (System Control Automation and Data Acquisition) is base to oil transporting in a safety conditions

That the greatest amount of data can be gathered, analyzed and acted upon in the shortest amount of time. Most operators now have some form of computer based monitoring system employing either commercially available or custom developed software to run the system. Programs can calculate the inventory of the line at any point in time and compare this with line measurement to provide independent cross checking of pipe line measurement data. These data can be transmitted by telemetry, radio, or telephone links to a central computer which continuously monitors the "health" of the line. The central operation room I now more likely o contain VDU's rather than banks of process indicators and enables he engineers to have rapid and accurate updates of conditions along the entire lie or network. By changing programs and subroutines, a large number of task and functions can be performed easily, simultaneously and cost effectively.

35

The many functions that can be performed by computer based systems include:
- Leak detection and location;
- Batch tracking of fluids;
- Pig tracking;
- On line flow compensation;
- Real time

Such systems have rapid response, but are dependent on the accuracy of the input data to be effective. Most computer based systems consists pf two major elements, namely:

1.a supervisory computer plus associated software;
2.a number of independent pipe line monitoring station plus data transmission equipment for communication with the computer.

The complex of the software, the number of input variables and the pipe line instrument selected, all influence the size and the choice of computer. Some systems are capable of running on the computer powerful micros such as the IBM, while the more complex pipe line networks will require a main frame machine, possibly communicating with satellite micros, installed close to the measurement stations. The overall design varies with each systems but should be underestimated, especially if SCADA systems are being development in-house (see section following).At the most basic level the computer system can take flowmeter inputs to perform gross balances along the pipe line. Such a method is very useful in identifying small but consistent loss of product from corrosion pits in the pipe wall. This will arise if the integrated flowmeter outputs diverge with time when the flow in the line is maintained constant. In the line flow rate varies with time then imbalances are more difficult to detect, since the meters at each end may have different characteristics or their outputs may vary nonlinearly with flow rate. A loss of product will be identified simply as the difference between the steady state inventory of the system and the instantaneous inlet and outlet flows.

Mathematically this is:
$$\Delta V = V_{in} - V_{out} - V_l \qquad (1)$$

Where $\Delta V$ is the leakage volume, $V_{in}$ and $V_{out}$ are the metered inlet and outlet flows, and $V_l$ is the inventory of the pipe line. This last term can be calculated as the average of the integrated flowmeter signal, but is dependent on variation in temperature, pressure and many other variables. It is use of computers that last term to be calculated on-line and as function of pipeline elevation, pipe material characteristics, process variation, and pipe line fitting loss data amongst the meant variables. Temperature and pressure variation change both the density of the fluid and the actual dimension of the line. Both will give rise to sources of errors in pipe line inventory if compensation is not included in the calculation. Figure 1 shows typical input data requirements for simple on line inventory calculation. In the other extreme, computer based systems can control entire networks with almost no need for the operators to intervene unless serious problems arise. Figure 3 shows example of pipeline monitoring systems.

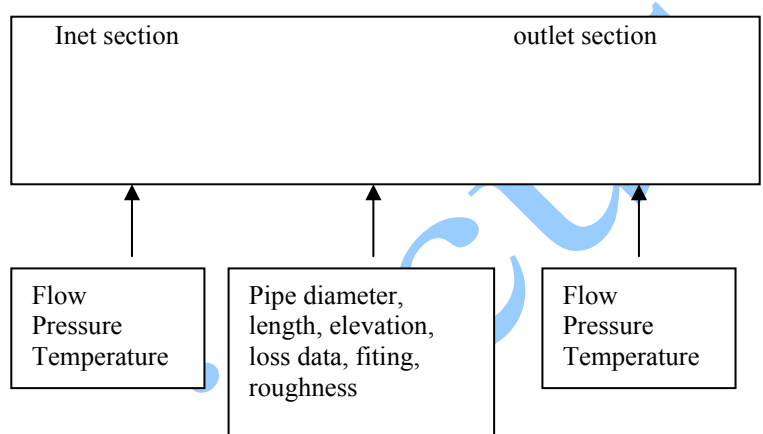

**Figure 1.** Input data for computer inventory calculation

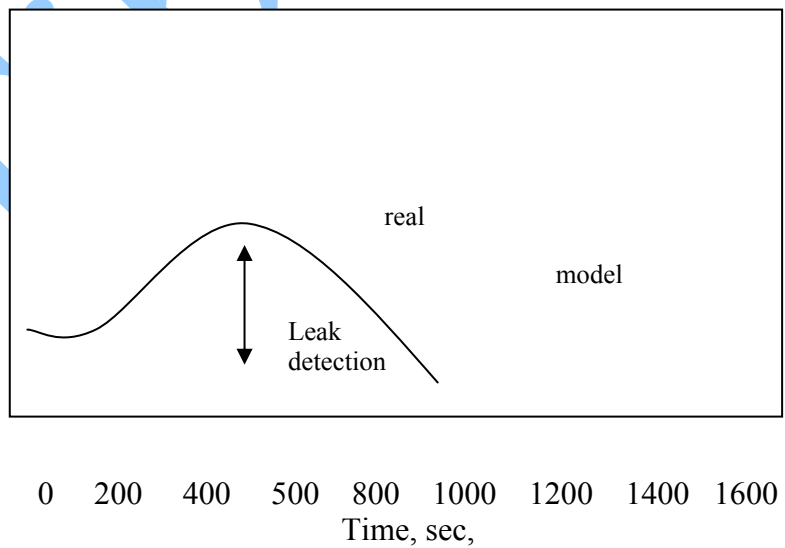

**Figure 2.** Real time modeling of pipe lines

The benefits from the addition of real time pipe line models cannot be overemphasized. Even when operating transients occur, some commercially available programs enable problem to be detected. An example is given in Figure 2. Here a pipe line leak as induced following pump failure and restart, and although the lie was operating under highly transient conditions, a material imbalance was detected. Real time modeling also allows density profiles to be calculated along the line. It is fairly common practice to have amounts of different fluids in the same pipe line separated by a pig or sphere. From the pressure drop/flow characteristics, the real time can track the progress of these batches of fluids along the pipe.

## 2   Leak Characterization

Leaks can occur from pinhole sized perforations caused by corrosion up to catastrophic pipeline failure due to manmade damage or natural causes such as an earthquake or a tsunami. Even relatively small holes in a high-pressure oil line can produce dangerous clouds of oil and gas. Leaks are actually very common and are classified as to the urgency of repair based on their potential danger. Typically they are classified into three groups, those that need repair in 24 to 48 hours, those which need to be repaired in 30 days, and those that don't need immediate repair, but should be monitored. Once gas escapes from the piping it can saturate the ground around the pipe and migrate along other conduits to other locations. The appearance of a rupture, leak or damage, which could cause a leak, usually generates an acoustic signal. During the crack initiation and early crack growth the steel pipe wall deformation creates a significant acoustic signal which can produce a transducer output ranging from several micro volts to several volts. The amplitude and frequency spectrum and the attenuation behavior are all a function of the pipe-wall material properties. If the damage causes a sudden leak then the associated rapid change in fluid pressure produces a pressure transient, often referred to as a burst signal. Once a leak is established the supersonic jet of escaping gas generates acoustic energy. These acoustic emissions are continuous and have a wide frequency spectrum (1kHz--1MkHz), the majority of which is confined to the moderately high frequency portion (175kHz – 750kHz).Passive acoustic leak detection in pipelines can make use of the vibration energy emitted by the straining or fracturing pipe wall material or by the acoustic energy associated with high pressure gas escaping through a perforated or ruptured wall. By properly interpreting the acoustic signature of these phenomena, it is possible to detect an infringement event along the pipeline. The challenge is to accurately isolate the acoustic signature of an infringement from the background noise within the pipeline environment such as pumping noise, flow turbulence noise, valve actuation, etc. Details of the infringement generated noise (acoustic signature) must be known as well as the details of the background noise within the pipe to enable separation between these two noises. A second challenge is to detect the acoustic signature far away from its source since the acoustic wave amplitudes are attenuated within the pipeline. The frequencies of the acoustic signatures of the structural fracturing of the pipe wall and the sound of the escaping gas can range well into the hundreds of kilohertz. Generally, the signal frequencies transported by the gas are lower and travel slower than those in the pipe wall. However, due to the intimate contact of the pipeline with the backfill material, the longitudinal transmission of the higher frequency components of acoustic energy within the wall material is highly damped and does not travel any significant distance from the location of the source of the signal. Damping in proportion to the square of the frequency impedes transmission of acoustic signals through gas. Viscous effects, wall damping effects, and molecular relaxation effects all contribute to the attenuation of the strength of the high frequency signal. Past acoustic studies have shown that while the acoustic signals of a pressurized fluid escaping through a leak may include a wide range of frequencies, only the relatively low frequencies are useful for practical leak detection methods due to the significant attenuation of the higher frequency components. Rocha states that acoustic frequencies on the order of 10Hz can propagate in a gas for distances on the order of 100 miles and gives the following approximation: the amplitude of the wave is related to the properties of the gas, the pressure at which the pipeline is being operated and the size of the leak. The local pressure drop due to the leak is given for a pipe without flow by:

$$\Delta p = 0.3 P_S (D_1/D_p)^2 \qquad (2)$$

where:  $\Delta p$ is the acoustic signal,
   $P_S$ is the static pressure in the pipe at the leak site,
   $D_1$ is the diameter of the leak hole,
   $D_p$ is the local diameter of the pipe.

The detectable acoustic pressure of a leak can be as small as 5 millibars (0.073 psi) in a pipeline with a static pressure of 69 bars (1000 psi).
This will require sophisticated noise cancellation techniques to increase the signal to noise ratio.

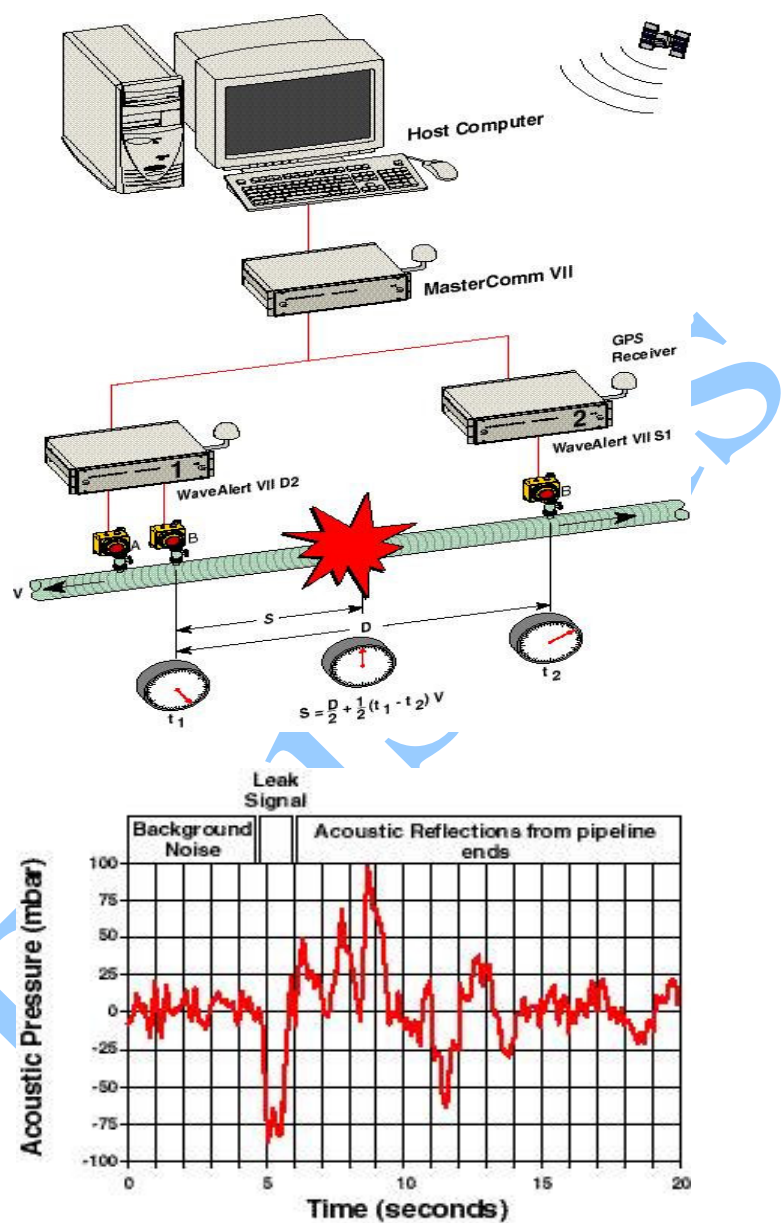

**Figure 3**. Pipeline monitoring systems

## 3   Wavelet Analysis in Signal Processing

The development of a sudden leakage initiates a pressure wave directly proportional to the size of the leak. When an underground pipeline is only accessible at the shut-off valves, which may be 60 km apart, then to locate the leak, the pressure pulse must be positively identifiable at both access points of the pipe. After traveling such a distance the signal to noise ratio may be too low for identification by conventional means.

The Fourier transform compares a signal with a set of sine and cosine functions. Each sine and cosine function oscillates at a different frequency. Hence, the Fourier transform indicates magnitudes of the signal at each individual frequency. On the other hand, the wavelet transform compares a signal with a set of short waveforms (called wavelets). Each wavelet has a different time duration (or scale). The shorter the time duration, the wider the frequency bandwidth, and vice versa. In mathematical jargon, the process of stretching or compressing the fundamental wavelet (usually called the mother wavelet) is named dilation. As wavelets get narrower and narrower, eventually they become impulse like functions (equivalent to wide frequency band). Consequently the wavelet transform is very powerful for detecting impulse like signals when a leak (theft) incident occurs, the oil pressure in the vicinity of the leakage point drops rapidly. Such a drop is presumably propagated in all directions along the pipeline. Consequently:
1.  Oil pressure decreases at both the inlet and outlet
2.  The oil flow rate at the outlet decreases, while the oil flow rate at the inlet increases. Based on the time difference of the pressure drops observed at the inlet and the outlet, Conceptually the leakage location can then be determined by   0.5* (length of pipeline +pressure wave velocity*time difference)

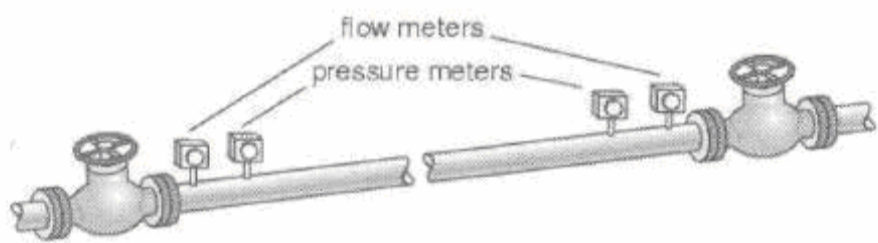

**Figure 4**.Oil pipe equipment

The difficulties in implementation of this application include:
1. Synchronization of all pressure and flow meters that are typically 60 km apart.
2. Variation of pressure wave velocity.

The pressure wave velocity is related to temperature and density of the medium, as well as the elasticity of the pipe material. To facilitate oil movement, the raw oil is often heated at each station, especially in cold weather. Due to the non-uniform temperature distribution, the pressure wave velocity is not constant. Consequently, the actual formula for estimating the location of the leak is much more involved than one might anticipate. Background noise as shown below, is high so that leakage associated rapid changes of pressure are often not noticed. For a typical pressure signal there is no obvious indication of the pressure wave. However, by applying the wavelet transform one can accurately determine the time of occurrence of the pressure drop.

Figure 5 depicts the signal after applying the wavelet transform. It shows signals recorded between 3:00 to 4:00 am on April 12, 2005. The upper plot shows the wavelet transform of the signal at the pipeline inlet, whereas the lower plot is wavelet transform of the signal at the pipeline outlet. As indicated in the wavelet transform domain, there were two incidents of leakage from the pipeline. The first one occurred between 3:00 and 3:18 am and the second between 3:24 and 3:44 am. For the second leakage, the time difference between the pressure drop at the pipeline inlet and the drop at the other end of the pipeline outlet was computed as 14.95 seconds. The leakage location was identified as 39,34 km south of Constanta Station. The actual point of theft was 39.29 km south of the station (less than 50 meters away!) Since the length of the pipeline is 61.48 km, the relative error was 0.26%.

**Conclusion**

The paper cannot do justice in such a short space, to the complex and diverse subject of computer pipeline systems. Such systems have been in operation in many forms all over the world, but it is only recently that environmental as well as economic factors have influenced their development.

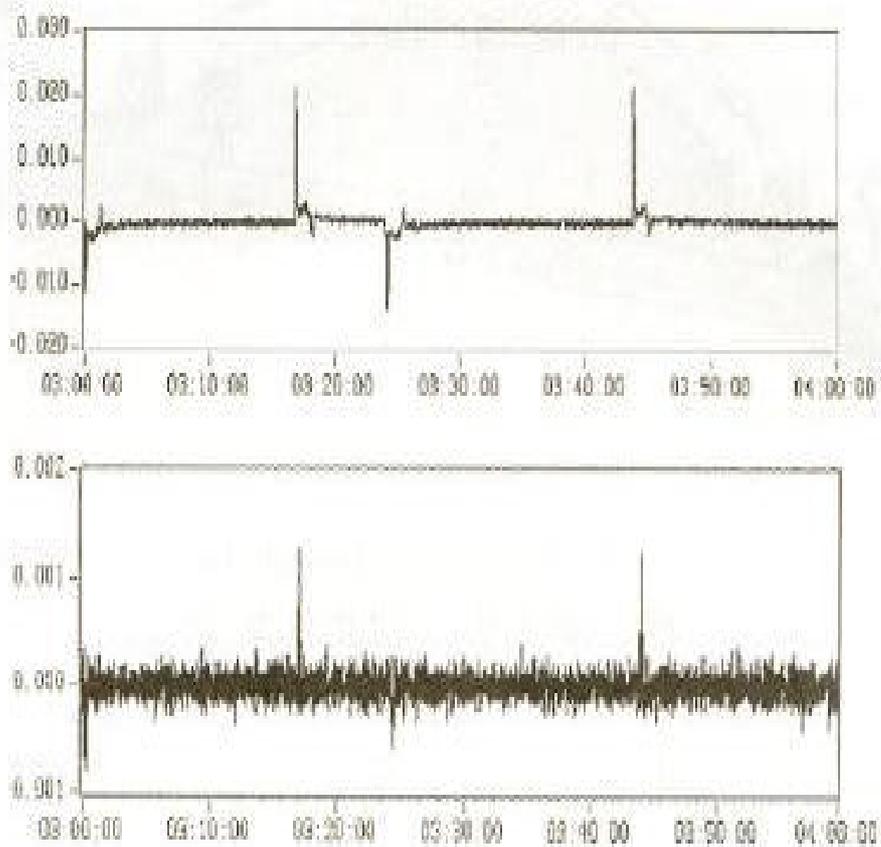

**Figure 5** - Acoustic detection of oil theft from a 60 km pipeline

**References**

[Tim05]   **Chis Timur**, „*Modern Pipe Line Monitoring Techniques*", First International Symposium Of Flow measurement and Control, Tokyo, Japan, 2005 .